\documentclass[10pt,a4paper]{article}
\usepackage{latexsym}
\usepackage{graphics}
\newtheorem{Def}{Definition}
\newtheorem{The}[Def]{Theorem}
\newtheorem{Pro}[Def]{Proposition}
\newtheorem{Lem}[Def]{Lemma}

\newtheorem{Rem}{Remark}
\newtheorem{Exa}{Example}

\newcommand{\N}{{\rm I\!N}}

\def\r#1{{\rm r}_{#1}}
\def\val#1{{\rm val}_{#1}}
\begin{document}
\title{Generalization of automatic sequences for numeration systems on a regular language}
\date{June 21, 1999}
\author{Michel Rigo\\
Institut de Math\'{e}matiques, Universit\'{e} de Li\`{e}ge, \\
Grande Traverse 12 (B 37), B-4000 Li\`{e}ge, Belgium.\\
{\tt M.Rigo@ulg.ac.be}}
\maketitle
\begin{abstract}
Let \(L\) be an infinite regular language on a totally ordered alphabet 
\((\Sigma,<)\). Feeding a finite deterministic automaton (with output)
with the words of \(L\) enumerated lexicographically  with respect 
to \(<\) leads to an infinite sequence over the output alphabet of the 
automaton. This process generalizes the concept of \(k\)-automatic sequence 
for abstract numeration systems on a regular language (instead of 
systems in base \(k\)). Here, I study the first properties of these sequences 
and their relations with numeration systems.   
\end{abstract}

\section{Introduction}
In \cite{LR}, P.~Lecomte and I have defined  a {\it numeration  system} 
as being a triple \(S=(L,\Sigma,<)\) where \(L\) is an infinite regular 
language over a totally ordered alphabet \((\Sigma,<)\).
The lexicographic ordering of \(L\) gives a one-to-one correspondence 
\(r_{S}\) between the set of the natural numbers \(\N\) and the language \(L\). 

For a given subset \(X\) of \(\N\), a question arise naturally. 
Is it possible to find a numeration system \(S\) such that 
\(\r{S} (X)\) is recognizable by finite automata ? (In this case, \(X\) 
is said to be {\it \(S\)-recognizable}.) For example, the set \(\{n^2:n\in 
\N\}\) is \(S\)-recognizable for some \(S\) and the arithmetic 
progressions \(p+q\N\) are \(S\)-recognizable for any \(S\). An interesting question 
is thus the following: is there a system \(S\) such that the set of primes 
is \(S\)-recognizable ?

To answer this question I show that a subset of \(\N\) is 
\(S\)-recognizable if and only if its characteristic sequence can be 
generated by an `automatic' method. The term automatic refers, as we 
shall see further, to a generalization of the \(k\)-automatic sequences 
for numeration systems on a regular language.

The \(k\)-automatic sequences are well-known and have been extensively 
studied since the 70's \cite{A2,D,Co,S1}. The construction of this kind 
of sequences is 
based on the representation of the integers in the base \(k\). For a given 
integer \(n\), one represents this number in base \(k\) using 
the greedy algorithm and obtains a word \([n]_{k}\) over the alphabet \(\{0,\ldots ,k-1\}\). 
Next one gives \([n]_{k}\) to a deterministic finite automaton with output 
and obtains the \(n^{th}\) term of a sequence which is said to be a \(k\)-automatic sequence.

These sequences have been already generalized in different ways \cite{A2}.
In particular, a method used by J. Shallit to generalize the \(k\)-automatic 
sequences is to consider some kind of linear numeration system instead of the 
standard numeration system with integer base \(k\) \cite{S1}. Two 
properties of the systems encountered in \cite{S1} are precisely that the set of all the 
representations is regular and that the lexicographic ordering is 
respected.

Here, instead of giving \([n]_{k}\) to a deterministic finite automaton with output, 
we feed it with \(\r{S}(n)\) to obtain an output which is the \(n^{th}\) 
term of an  
{\it \(S\)-automatic sequence} for a numeration system \(S\). Having thus 
introduced the concept of \(S\)-automatic 
sequences, we can follow two paths. Learn 
their intrinsic properties but also use them as a tool 
to check if a subset of \(\N\) is \(S\)-recognizable.

Our article has the following articulation.
In the first section, we recall 
some definitions and we introduce a teaching example which could be 
very instructive for the reader not familiar with automatic sequences. 
In the second section, we adapt the classical 
results concerning the fiber and the kernel of an automatic sequence. 

Initially, A. Cobham showed the equivalence between the \(k\)-automatic 
sequences and the sequences obtained by iterating a uniform morphism 
(also called uniform tag system \cite{Co}). In the third section,
we show that an \(S\)-automatic sequence is always generated by a 
substitution (i.e., an iterated non-uniform morphism followed by one 
application of another morphism). From this, we deduce that the number of distinct 
factors of length \(l\) in an \(S\)-automatic sequence is in \(O(l^2)\). We also show how to 
construct \(S\)-automatic sequences with at least the same complexity that 
infinite words obtained by iterated morphisms.

In the last section, we will be able to show that for any numeration 
system \(S\), the set of primes is never \(S\)-recognizable. We 
use the fact that to be \(S\)-recognizable, the characteristic 
sequence of the set must be generated by a substitution. Hence we use some 
results of C.~Mauduit about the density of the infinite words 
obtained by substitution \cite{M1,M2}. 
\section{Basic definitions and notations}

In this paper, capital greek letters represent finite alphabet. We denote 
by \(\Sigma^*\) the set of 
the words over \(\Sigma\) (\(\varepsilon\) is the empty word) and by \(\Sigma^\omega\) the set of the 
infinite words over \(\Sigma\). If \(K\) is a set then \(\# K\) denotes the cardinality of \(K\) and if \(w\) is 
a string then \(|w|\) denotes the length of \(w\). For \(1\le i\le |w|\), 
\(w_{i}\) is the \(i^{th}\) letter of \(w\). The same notation holds 
for infinite words, in this case \(i\in \N\setminus\{0\}\).

First, recall some definitions about the numeration systems we are dealing with. For more about these 
systems see \cite{LR}.

\begin{Def}{\rm
A {\it numeration system} \(S\) is a triple \((L,\Sigma,<)\) where \(L\) is an infinite regular language over the 
totally ordered alphabet \((\Sigma,<)\).

For each \(n\in\N\), \(\r{S}(n)\) denotes the \((n+1)^{th}\)  word of \(L\)  with respect to the
lexicographic ordering and is called the {\it \(S\)-representation} of \(n\).

Remark that the map \(\r{S}: \N \to L\) is an increasing
bijection. For \(w\in L\), we set \(\val{S} (w)=\r{S}^{-1} (w) \). We call \(\val{S}(w)\) the {\it numerical value} of \(w\).
}\end{Def}

Examples of such systems are the numeration systems defined by a recurrence relation whose
characteristic polynomial is the minimum polynomial of a Pisot number 
\cite{BH}. (Indeed, with this hypothesis, the set of representations 
of the integers is a regular language.) The standard numeration systems with 
integer base and also the Fibonacci system belong to this class.

\begin{Def}{\rm 
Let \(S\) be a numeration system. A subset \(X\) of \(\N\) is {\it \(S\)-recognizable} if \(\r{S} (X)\) 
is recognizable by finite automata.
}\end{Def}

Let us introduce the concept of \(S\)-automatic sequence which naturally generalizes the \(k\)-automatic sequences based 
on the representation of the integers in base \(k\). For more about \(k\)-automatic sequences see for instance \cite{A2,D}.

\begin{Def}{\rm A {\it deterministic finite automaton with output} (DFAO) \(M\) is a \(6\)-uple 
\((K,s,\Sigma,\delta,\Delta,\tau)\) where \(K\) is the finite set of the states, \(s\) is the start state, 
\(\Sigma\) is the input alphabet, \(\delta: K\times \Sigma \to K\) is the transition function, 
\(\Delta\) is the output alphabet and \(\tau: K \to \Delta\) is the output function.
}\end{Def}

\begin{Def}{\rm Let \(S=(L,\Sigma ,<)\) be a numeration system. A sequence \(u\in \Delta^\omega\) is 
{\it \(S\)-automatic} if there exists a DFAO \(M=(K,s,\Sigma,\delta,\Delta,\tau)\) such that for all 
\(n \in \N\), 
\[u_{n+1}=\tau(\delta(s,\r{S} (n))).\]
If the context is clear, we write \(\tau(w)\) in place of \(\tau(\delta(s,w))\).
}\end{Def}

\begin{Rem}{\rm A subset \(X\subset \N\) is \(S\)-recognizable if and only if its characteristic sequence 
\(\chi_X \in \{ 0,1 \}^\omega\) is \(S\)-automatic.
}\end{Rem}

In the following we will often encounter two more `classical' ways of 
obtaining infinite sequences.

\begin{Def}{\rm Let \(\varphi:\Sigma\to\Sigma^*\) be a morphism of 
monoid such that for some \(\sigma \in \Sigma\), \(\varphi(\sigma) \in 
\sigma \Sigma^*\). The word \(u_{\varphi}=\varphi^\omega (\sigma)\) is a fixed 
point of \(\varphi\) and we say that \(u_{\varphi}\) is {\it generated by 
an iterated morphism}.

A morphism is {\it uniform} if \(|\varphi(\sigma_{1})|=\ldots 
=|\varphi(\sigma_{n})|\), \(\Sigma=\{\sigma_{1},\ldots ,\sigma_{n}\}\).
}\end{Def}

\begin{Def}{\rm A {\it substitution} \(T\) is a triple 
\((\varphi,h,c)\) such that \(\varphi:\Sigma\to\Sigma^*\) and 
\(h:\Sigma\to\Delta^*\) are morphisms of monoids. Moreover \(c 
\in \Sigma\), \(\varphi(c) \in c \Sigma^*\) and for any \(\sigma \in 
\Sigma\), \(h(\sigma)=\varepsilon\) or \(h(\sigma) \in \Delta\) (\(h\) 
is said to be a {\it weak coding}). We 
said that the word \(u_{T}=h(\varphi^\omega(c))\) over 
\(\Delta\) is {\it generated by the substitution} \(T\).

If \(h(\sigma)=\varepsilon\) for some \(\sigma\) then \(h\) is said to 
be {\it erasing} otherwise \(h\) is said to be {\it non-erasing}.
}\end{Def}

\subsection{A teaching example}
We consider the numeration system \(S=(a^* b^*,\{ a,b \},a<b)\), the 
alphabets \(\Sigma=\{ a,b\}\), \(\Delta=\{0,1,2,3\}\) and the 
following DFAO
\begin{center}
\includegraphics{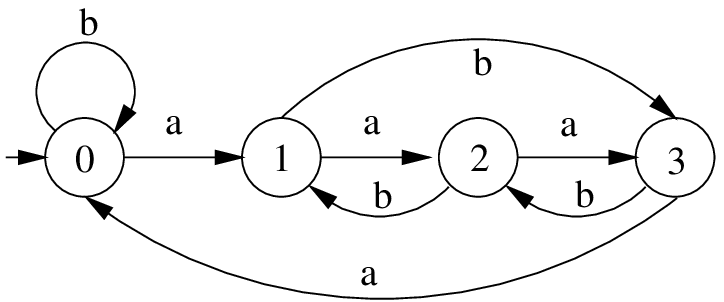}
\end{center}
As usual the start state is indicated by an unlabeled arrow. The first words of \(a^* b^*\) are 
\[\varepsilon,a,b,aa,ab,bb,aaa,aab,abb,bbb,\ldots\]
and thus feeding 
the automaton with these words we obtain the first terms of the sequence 
\(u\in \Delta^\omega\),
\[u=01023031200231010123023031203120231002310123010123 \ldots .\]

\begin{Rem}{\rm The sequence \(u\) is not ultimately periodic. One can observe that the distance 
between two occurrences of the block `\(00\)' is not bounded. Indeed, 
\begin{equation}\label{eq1}
\tau(w)=0 \Leftrightarrow \exists r,s \in \N : w=a^{4r} b^s 
\end{equation}
thus a block `\(00\)' comes from two consecutive words \(b^{4r-1}\) and \(a^{4r}\), \(r\ge 1\) and the 
number of words of length \(n\) in \(a^* b^*\) is \(n+1\), \(n \in \N\). 
}\end{Rem}

\begin{Rem}{\rm The sequence \(u\) is not generated by an iterated 
morphism \(\varphi\). First observe that
\[
\tau(w)=\left\{\begin{array}{l}1\\ 2\\ 3\end{array} \right\} \Leftrightarrow 
\exists r,s \in \N : w= 
\left\{\begin{array}{l}
a^{4r+1} b^{3s},\ a^{4r+2} b^{3s+1},\ a^{4r+3} b^{3s+2} \\
a^{4r+1} b^{3s+2},\  a^{4r+2} b^{3s},\ a^{4r+3} b^{3s+1}  \\
a^{4r+1} b^{3s+1},\  a^{4r+2} b^{3s+2},\ a^{4r+3} b^{3s}.
\end{array}\right.
\]
Suppose that there exists a morphism \(\varphi\) such that 
\(u = \lim_{n\to +\infty} \varphi^n (0)\).

1) If \(\varphi (0) \in 0102\Delta^*\) then the block `\(0102\)' must 
appear at least twice in \(u\) since `\(0\)' appears twice in \(u\). 
If the first `\(0\)' of the block is obtained from a word \(a^{4r} b^s\) with 
\(r\ge 1\) then the second `\(0\)' is obtained from \(a^{4r-2} b^{s+2}\) 
which leads to a contradiction in view of (\ref{eq1}). If the first `\(0\)' is obtained from \(b^s\) with 
\(s\ge 1\) then the second `\(0\)' come from  \(a^s b\) and we have 
\(s=4t\). The `\(2\)' is obtained from \(a^{4t-1} b^2\) 
which also leads to a contradiction.

2) If \(\varphi (0)=01\) then in view of the first terms of 
\(u\), \(\varphi (1) \in 023031200231\Delta^* \). We show that 
`\(023031200\)' appears only once in \(u\). Suppose that we can 
find another block of this kind. Thus the last two `\(0\)' come from  
words \(b^{4r-1}\) and \(a^{4r}\) with \(r\ge 2\). Since we consider 
all the words of \(a^*b^*\) lexicographically ordered, the first `\(0\)' 
of the block come from \(a^{7} b^{4r-8}\) which is in contradiction 
with (\ref{eq1}).

3) If \(\varphi (0)=010\) then \(\varphi (1) \in 23031200231\Delta^* \) 
and \(\varphi (010) \in 01023031200\Delta^* \). The block `\(010\)' 
appears at least twice in \(u \) but we know that `\(023031200\)' appears only once. 

We shall see further that \(u\) is generated by a substitution.
}\end{Rem}
\section{First results about \(S\)-automatic sequences}
Some classical results about \(k\)-automatic sequences can be easily 
restated \cite{Co,D}.

\begin{Def}{\rm Let \(a\in \Delta\) and \(S=(L,\Sigma ,<)\) , the {\it \(S\)-fiber} \({\cal F}_S (u,a)\) 
of a sequence \(u \in \Delta^{\omega}\) is defined as follows
\[{\cal F}_S (u,a) = \{ \r{S} (n) \, : \, u_n = a \}.\]
}\end{Def}

\begin{The} Let \(u\) be an infinite sequence over \(\Delta\) and \(S=(L,\Sigma ,<)\). The sequence \(u\) 
is \(S\)-automatic if and only if for all \(a \in \Delta\), \({\cal F}_S (u,a)\) is a regular subset 
of \(L\).
\end{The}
{\it Proof.} If \(u\) is \(S\)-automatic then we have a DFAO 
\(M=(K,s,\Sigma,\delta,\Delta,\tau)\) which is used to generate \(u\). 
Let \(L(M')\) be the language recognized by the DFA 
\(M'=(K,s,\Sigma,\delta,F)\) where the set of final 
states \(F\) only contains the states \(k\) such that \(\tau(k)=a\). 
Therefore \({\cal F}_S (u,a)\) is regular since it is the 
intersection of the two regular sets \(L(M')\) and \(L\).

The condition is sufficient. Let \(\Delta =\{a_1,\ldots ,a_n \}\). 
Remark that if \(i\neq j\), \({\cal F}_S (u,a_i) 
\cap {\cal F}_S (u,a_j) = \emptyset\) and \(L=\cup_{i=1}^n {\cal F}_S 
(u,a_i) \). For all \(i=1,\ldots ,n\),  \({\cal F}_S (u,a_i) \) is 
accepted by a DFA \(M_i=(K_i,s_i,\Sigma,\delta_i,F_i)\). From these 
automata we construct a DFAO \(M=(K,s,\Sigma,\delta,\Delta,\tau)\) to 
generate \(u\) using the 
numeration system \(S\). The set \(K\) is \(K_1\times \ldots \times K_n\), 
the initial state is \((s_1,\ldots,s_n)\). For all states 
\((q_1,\ldots,q_n) \in K\) and for all \(\sigma \in \Sigma\), 
\(\delta((q_1,\ldots,q_n),\sigma)=(\delta_1(q_1,\sigma),\ldots 
,\delta_n(q_n,\sigma))\). If there is a unique \(i\) such that 
\(q_i \in F_i\) then \(\tau((q_1,\ldots ,q_n))=a_i\) otherwise the 
state cannot be reached by a word of \(L\) and the output is not 
important. The sequence \(u\) is obtained from \(S\) and the DFAO 
\(M\) thus \(u\) is \(S\)-automatic. \(\Box\)

\medskip
\noindent
The notion of \(k\)-kernel of a \(k\)-automatic sequence can be transposed as follows.

\begin{Def}{\rm Let \(S=(L,\Sigma ,<)\) and \(u\) be an infinite sequence. For each \(w \in \Sigma^*\), we set 
\({\cal K}_w = \{ v \in L \, |\, \exists z\in \Sigma^* : v=wz\}.\) One can 
enumerate \({\cal K}_w\) lexicographically with respect to \(<\), \({\cal 
K}_w=\{ wz_0 < wz_1 < \ldots \}\). 
Thus for each \(w \in \Sigma^*\), one can construct the subsequence \(n\mapsto 
u_{\val{S} (wz_n)}\) (remark that the subsequence can be finite or even empty).
}\end{Def}

\begin{The} Let \(S=(L,\Sigma,<)\). 
A sequence \(u\in \Delta^\omega\) is \(S\)-automatic if and only if 
\(\{ n\mapsto u_{\val{S} (wz_n)} \, : \, w \in \Sigma^* \}\) 
is finite.
\end{The}
{\it Proof.} If \(u\) is \(S\)-automatic, we have a DFAO 
\(M=(K,s,\Sigma,\delta,\Delta,\tau)\) used to generate \(u\) 
and we define the equivalence relation \(\sim_1\) over \(\Sigma^*\)  
by \(x \sim_1 y\) if and only if \(\delta(s,x) = \delta(s,y)\). In 
the same way, the minimal automaton of \(L\) provides an equivalence 
relation \(\sim_2\). The two relations have a finite index thus the 
relation \(\sim_{1,2}\)  given by \(x \sim_{1,2} y\) if and only if \(x \sim_1 y\) and \(x \sim_2 
y\) has also a finite index. Remark that each class of \(\sim_{1,2}\) 
gives one of the sequences \(n\mapsto u_{\val{S} (wz_n)}\). Indeed, 
\(x\sim_{2}y\) implies that \(\{z\in\Sigma^* : xz\in L\}=\{z\in\Sigma^* : 
yz\in L\}\) thus \({\cal K}_{x}=\{xz_{0}<xz_{1} <\ldots \}\) and 
\({\cal K}_{y}=\{yz_{0}<yz_{1} <\ldots \}\) with the same 
\(z_{0},z_{1},\ldots\).

The condition is sufficient. We show how to construct a DFAO. The 
states are the subsequences \(q_w=(n\mapsto u_{\val{S} (wz_n)})\).
The initial state is \(q_\varepsilon\) (i.e., the subsequence obtained from 
the empty word). The transition function \(\delta\) is given by 
\(\delta(q_w,\sigma)=q_{w\sigma}\) and the output function \(\tau\) 
is given by \(\tau(q_w)=u_{\val{S}(w)}\). \(\Box\)

\section{Complexity of \(S\)-automatic sequences}
The complexity function \(p_u\) of an infinite sequence \(u\) maps \(n\in \N\) to the number \(p_u(n)\) of distinct factors of length \(n\) which 
occur at least once in \(u\). In this section, we will show that the 
complexity of an \(S\)-automatic sequence is in \(O(n^2)\) as a 
consequence that every \(S\)-automatic sequence is generated by a 
substitution.

Recall that an infinite word \(w\) generated by iterated morphism has a complexity 
such that
\[ c_{1} f(n) \le p_{w}(n) \le c_{2} f(n) \]
where \(f(n)\) is one of the following functions
\(1\), \(n\), \(n \log \log n\), \(n \log n\) or  \(n^2\) \cite{Pa}. 
For a survey on the complexity function, see for example \cite{A1}.

The next remark shows that an \(S\)-automatic sequence can 
reach at least the same complexity as a word generated by morphism.

\begin{Rem}\label{RemComplex}{\rm For every infinite word \(w\) generated 
by an iterated morphism \(\varphi\) 
over an alphabet \(\Delta\) we can construct an \(S\)-automatic 
sequence \(u\) such that \(\forall n\in \N,\) \(p_w (n) \le p_u (n)\).

We show how to proceed on the following example,
\[ \Delta=\{ 0,1 \},\ 
\varphi : \left\{ \begin{array}{l}
0\mapsto 0101 \\
1\mapsto 11.
\end{array}\right.
\]
It is well-known that \(w=\varphi^\omega (0)\) is such that \(p_w\) is 
of complexity \(O (n \log \log n)\) \cite{Pa}. To the morphism \(\varphi\), we associate a finite automaton \(M\) (if 
the morphism is not uniform then \(M\) is not deterministic). 
The set of states is \(\Delta\), all the states are final and the 
transition function \(\delta\) is obtained 
by reading the productions of \(\varphi\) from left to right. For 
this purpose, we introduce a new ordered alphabet 
\(\Sigma\) such that \(\# 
\Sigma= \sup_{x\in \Delta} |\varphi (x)|\). Here, \(0\) gives the 
initial state (for we consider the word \(\varphi^\omega (0)\)) and 
\(1\) the other state. Thus with \(\Sigma =\{a<b<c<d\}\), we have 
\(\delta(0,a)=[\varphi(0)]_1 = 0\), \(\delta(0,b)=[\varphi(0)]_2= 
1\), \(\ldots \) and \( M \) is then 
\begin{center}
\includegraphics{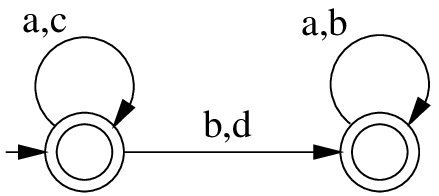}
\end{center}
As is customary, the final states are denoted by double circles. 
The language accepted by \( M \) is \(L= \{a,c\}^* \{b,d\} \{a,b\}^* \cup 
\{a,c\}^* \). The numeration system \(S\) is thus 
\((L,\Sigma,a<b<c<d)\).  This kind of construction can also be found 
in \cite{Ma}. Now from \(M\) we simply construct a 
DFAO \(M'\)
\begin{center}
\includegraphics{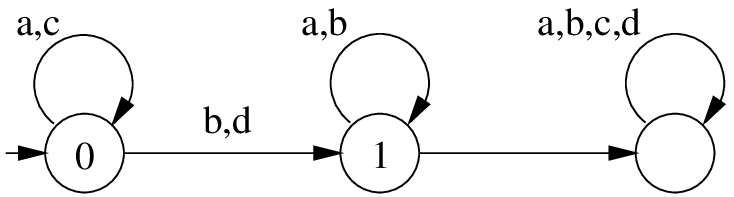}
\end{center}
The way we find the output can be easily understood. The third state 
can have any output for this state is never reached with a word 
belonging to \(L\). 
One remarks that the \(S\)-automatic sequence obtained with \(M'\) and 
\(S\) is 
\[ u=\varphi (0) \varphi^2 (0) \varphi^3 (0) \ldots \]
and thus every factor of \(w=\varphi^\omega (0) \) belongs to \(u\).
}\end{Rem}

We now show that every \(S\)-automatic sequence is generated by a 
substitution.

\begin{Lem}\label{SEtat} Let \(\Sigma=\{\sigma_{1}<\ldots <\sigma_{n}\}\), 
\(M=(K,s,\Sigma,\delta,F)\) be a DFA and \(\alpha \not\in K\). The 
morphism \(\varphi_{M} : K \cup \{\alpha\} \to (K \cup \{\alpha\})^*\) 
defined by
\[
\left\{\begin{array}{l}
\alpha  \mapsto  \alpha s \cr
k  \mapsto  \delta(k,\sigma_{1}) \ldots \delta(k,\sigma_{n}) ,\ k \in K 
\cr
\end{array}\right.
\]
produces the sequence \(u_{\varphi}\) of the states reached by the 
words of \(\Sigma^*\) i.e., \(\forall i\in \N\setminus\{0\}\), \(u_{i+1}=\delta(s,w_{i})\) where \(w_{i}\) 
is the \(i^{th}\) element of \((\Sigma^*,<)\). 
\end{Lem}
{\it Proof.} One can check easily by constructing \(\varphi(\alpha)\), 
\(\varphi^2(\alpha)\), \(\varphi^3(\alpha)\) (which are prefixes of \(u_{\varphi}\)) that \(u_{\varphi}\) 
satisfies the property. \(\Box\)

\begin{Pro}\label{Sub} Every \(S\)-automatic sequence is generated by a 
substitution. 
\end{Pro}
{\it Proof.} Let \(S=(L,\Sigma,<)\), \(M_{L}=(K,s,\Sigma,\delta,F)\) 
be a DFA accepting \(L\) and \(u\) be an \(S\)-automatic sequence obtained with the DFAO 
\({\cal M}=(K',s',\Sigma,\delta',\Delta,\tau)\). From these two 
automata, we construct the product automaton \(M=(K\times K', (s,s'), 
\Sigma, \nu)\) where 
\( \nu ((k,k'),\sigma)=( \delta (k,\sigma),\delta' (k',\sigma )) \). We do not 
give explicitly the final states of \(M\). By Lemma \ref{SEtat}, we associate 
to this automaton a morphism \(\varphi_{M} : (K\times K') \cup \{\alpha\} 
\to ((K\times K') \cup \{\alpha\})^*\). To conclude the proof, we 
construct the erasing morphism \(h : (K\times K') \cup 
\{\alpha\} \to \Delta^*\) defined by
\[
\left\{\begin{array}{rll}
h(\alpha)&=\varepsilon \cr
h((k,k'))&=\varepsilon & \ {\rm if}\ k \not\in F \cr
 & =\tau(k') &\ {\rm otherwise}. \cr
\end{array}\right.
\]
Indeed, \(\varphi^\omega(\alpha)\) is the sequence of the states 
reached by the words of \(\Sigma^*\) in \(M\) but we are only 
interested in the words belonging to \(L\) and in the corresponding 
output of \({\cal M}\). Thus \(u\) is generated by 
\((\varphi_{M},h,\alpha)\). \(\Box\)
\bigskip

Dealing with erasing morphisms whenever one wants to determine the complexity 
function of a sequence is painful. So the next lemma permits to get rid of 
erasing morphisms.
\begin{Lem}\label{Efface} {\rm \cite{A3}}
If \(f\) and \(g\) are arbitrary morphisms with \(f(g^\omega (a))\) 
an infinite word, then there exists a non-erasing morphism \(k\) and a 
coding \(h\) (i.e., a letter-to-letter morphism \(h\)) such 
that \(f(g^\omega (a))=h(k^\omega (a))\). \(\Box\)
\end{Lem}

\begin{The} The complexity of an automatic sequence is in \(O(n^2)\). Moreover, there exists an 
automatic sequence \(v\) and a positive constant \(d'\) such 
that \(\forall n>0\), \(p_{v} (n)\ge d' n^2\).
\end{The}
{\it Proof.} Let \(u\) be an \(S\)-automatic sequence. By Proposition 
\ref{Sub}, \(u\) is generated by a substitution \((\varphi,h,\alpha)\) 
and by Lemma \ref{Efface} we can suppose that \(h\) is non-erasing. 
The word \(u_{\varphi}=\varphi^\omega (\alpha)\) is generated by an iterated 
morphism and thus \(p_{u_{\varphi}} (n) \le d \, n^2\). 
To conclude, since \(u=h(u_{\varphi})\), recall that if \(v\), \(w\)
are two infinite words and if \(h\) is a non-erasing 
morphism such that \(h(v)=w\) then there exist positive 
constants \(a\), \(b\) such that \(p_{w} (n) \le a\, p_{v}(n+b)\) \cite{Pa}. 

2) We show that there exist a language \(L\) over an ordered alphabet 
and a DFAO such that the corresponding 
automatic sequence \(v\) has a complexity function \(p_{v} (n)\ge d' 
n^2\). 

The morphism  
\[
\varphi:\left\{\begin{array}{l}
0\mapsto 01\cr
1\mapsto 12\cr
2\mapsto 2\cr \end{array}\right.
\]
generates the word \(w=\varphi^\omega (0)\). Since \(2\) is a bounded 
letter (i.e., \(|\varphi^n (2)|\) is bounded) and \(2^n\) is a factor of \(w\) for an arbitrary \(n\), there 
exists a positive constant \(d'\) such that \(p_{w} (n)\ge d'\, n^2\) 
(see \cite{Pa}). Using the 
same technique as in  Remark \ref{RemComplex}, we construct an 
\(S\)-automatic sequence \(v\) such that \(p_{v} (n)\ge p_{w} (n)\). 
One find easily that the regular language used in the numeration 
system \(S\) is \(L=a^* \cup a^*ba^* \cup a^*ba^*ba^* \). \(\Box\)
\medskip

To conclude this section, we refine in a very simple way 
Proposition \ref{Sub} to give a 
characterization of the \(S\)-automatic sequences.

Let \(T=(\varphi,h,c)\) and \(T'=(\varphi',h',c')\) be two 
substitutions such that \(\varphi:\Sigma\to\Sigma^*\), 
\(h:\Sigma\to\Delta^*\),\(\varphi':\Sigma'\to\Sigma'^*\) and  
\(h':\Sigma'\to\Delta'^*\). A morphism of substitutions \(m:T\to T'\) is a 
surjective morphism \(m:\Sigma\cup\Delta\to \Sigma'\cup\Delta'\) such that
\begin{enumerate}
\item \(m(c)=c'\), \(m(\Sigma)=\Sigma'\), \(m(\Delta)=\Delta'\)
\item \(m(\varphi(\sigma))=\varphi'(m(\sigma))\), \(\forall \sigma \in \Sigma\)
\item \(m(h(\sigma))=h'(m(\sigma))\), \(\forall \sigma \in \Sigma\).
\end{enumerate}

For a regular language \(L\) on the totally ordered alphabet 
\((\Sigma,<)\) and for a DFAO \(M=(K,s,\Sigma,\delta,\Delta,\tau)\), one 
can construct the {\it canonical substitution \(T_{(L,<,M)}\)} by 
proceeding in  the same way as in Proposition \ref{Sub} with 
\(M_{L}\) equals to the 
minimal automaton of \(L\) and the DFAO \({\cal M}\) equals to a 
reduced and accessible copy of \(M\).

To reduce \(M\), one have to merge the states \(p\), \(q\) 
such that for all \(w\in 
\Sigma^*\), \(\tau (\delta (p,w))=\tau (\delta (q,w))\).

\begin{Def}{\rm A substitution \(T\) is an 
{\it \((L,<,M)\)-substitution} if there exists a morphism \(m:T\to 
T_{(L,<,M)}\). This kind of construction has already been introduced 
in \cite{BH} for linear numeration systems based on a Pisot number.
}\end{Def}

The next theorem is obvious and we state it without proof.

\begin{The} Let \(S=(L,\Sigma,<)\). The sequence \(u\in 
\Delta^\omega\) is \(S\)-automatic if and only if \(u\) is generated 
by a \((L,<,M)\)-substitution for some DFAO \(M\). \(\Box\)
\end{The}
\section{Application to \(S\)-recognizable sets of integers}
Proposition \ref{Sub} gives a necessary condition for a set \(X\) of 
integers to be \(S\)-recognizable. The characteristic sequence 
\(\chi_{X}\in \{0,1\}^\omega\) has to be generated by a 
substitution. Thus this proposition can be used as an interesting tool 
to show that a subset of \(\N\) is not \(S\)-recognizable for any 
numeration system \(S\).

In the following \({\cal P}\) is the set of primes and \(\chi_{\cal 
P}\) is its characteristic sequence. We show that \({\cal P}\) is 
never \(S\)-recognizable but first we construct by hand a subset of 
\(\N\) which cannot be \(S\)-recognizable for its characteristic 
sequence is too complex.

\begin{Exa}{\rm For \(n\ge 3\), consider the \(\pmatrix{n\cr 3}\) 
words belonging to \(\{0,1\}^n\) which contains exactly three 
`\(1\)' and concatenate these words lexicographically ordered to 
obtain the word \(w_{n-3}\). To conclude consider the infinite word 
\[w=w_{0}w_{1}w_{2}\ldots = 
\underbrace{111}_{w_{0}} \underbrace{0111\,1011\,1101\,1110}_{w_{1}} 
\underbrace{00111\,01011\,\ldots}_{w_{2}} \ldots \]
By construction, it is obvious that for all positive constants \(C\), 
there exists \(n_{0}\) such that \(\forall n\ge n_{0} : p_{w}(n)>Cn^2\). 
Thus \(w\) cannot be generated by a substitution and the corresponding 
subset \(W\) such that \(\chi_{W}=w\), 
\[W=\{0,1,2,4,5,6,7,9,10,11,12,14,15,16,17,21,22,23,25,27,28,\ldots \},\]
is never \(S\)-recognizable.
}\end{Exa}

\begin{Pro} For any numeration system \(S\), \({\cal P}\) is not 
\(S\)-recognizable.
\end{Pro}
{\it Proof.} In \cite{M1,M2}, C.~Mauduit shows using some density 
arguments that \(\chi_{\cal P} \in \{0,1\}^\omega\) is not 
generated by a substitution \((\varphi,h,\alpha)\) where \(h\) sends 
all the letters on \(0\) except one. A slight adaptation of the proof 
leads to the conclusion for any letter-to-letter morphism \(h\). \(\Box\)
\section{Acknowledgments}
The author would like to warmly thank J.-P.~Allouche for pointing 
out very useful references and for some fruitful comments and also 
P.~Lecomte for his support and for fruitful conversations.

\end{document}